\documentclass[times,final,3p]{elsarticle}

\usepackage{amssymb}
\usepackage{latexsym}
\usepackage{amsmath}
\usepackage{dsfont}
\usepackage{natbib}
\usepackage{color}
\usepackage{amssymb}

\journal{Physica Scripta}

\begin{document}

\begin{frontmatter}

\title{Efficient algebraic solution for a time-dependent quantum harmonic oscillator}

\author[1]{Daniel M. Tibaduiza\corref{cor1}}
\cortext[cor1]{Corresponding author: 
  Tel.: +55-21-98084-9366;  }
\ead{danielmartinezt@gmail.com}
\author[1]{Luis Pires}
\author[2]{Andreson L.~C.~Rego}
\author[1]{Daniela Szilard}
\author[1]{Carlos Zarro}
\author[1]{Carlos Farina}

\address[1]{Instituto de F\'\i sica - Universidade Federal do Rio de Janeiro, Av. Athos da Silveira Ramos 149--Centro de Tecnologia, Rio de Janeiro 21941-972, Brazil.}
\address[2]{Instituto de Aplica\c{c}\~{a}o Fernando Rodrigues da Silveira - Universidade do Estado do Rio de Janeiro, Rua Santa Alexandrina 288, Rio de Janeiro 20261-232, Brazil.}

\begin{abstract}
Using operator ordering techniques based on BCH-like relations of the \textit{su}(1,1) Lie algebra and a time-splitting approach, 
we present an alternative method of solving the dynamics of a time-dependent quantum harmonic oscillator for any initial state. 
We find an iterative analytical solution given by simple recurrence relations that are 
very well suited for numerical calculations. We use our solution to reproduce and analyse some results from literature in order to prove 
the usefulness of the method and, based on these references, we discuss efficiency in squeezing, 
when comparing the parametric resonance modulation and the Janszky-Adam scheme.
\end{abstract}

\end{frontmatter}

\section{Introduction}

The time-dependent quantum harmonic oscillator (TDHO) is an important system in several branches of physics and has been a source of novel concepts for the past seven decades.  It is a natural scenario for the study of many important topics as time-dependent hamiltonians, 
foundations of quantum mechanics, mathematical theorems and the increasingly important squeezed states. Such states appear, for instance, 
in quantum optics \cite{WALLS-1983, LOUDON-1987, WU-1987, TEICH-1989, PERINA-1991, DODONOV-2002, DODONOV-2003, dutt-2015, SCHNABEL-2017, Raffa2019}, 
in cosmology \cite{GRISHCHUK-1990, Lo-ST-1991, GRISHCHUK-1993, ALBRECHT-1994, HU-1994, EINHORN-2003,KIEFER-2007}, and in some approaches to the dynamical Casimir effect, particularly, those based on analogue models \cite{DODONOV-2005, JOHANSSON-2009,JOHANSSON-2010,DODONOV-2010,WILSON-2011,FUJII-2011,LAHTEENMAKI-2013,FELICETTI-2014}.
The main property of squeezed states is that they provide variances of certain quadratures smaller than the value associated to coherent states \cite{BO-STURE-1985, WODKIEWICZ-1985, GAZEAU-2009}, enhancing the sensitivity of several systems \cite{VAHLBRUCH-2007, GIOVANNETTI-2011}. 
Some remarkable examples are in telecommunications \cite{SLAVIK-2010, Fedorov2016, Pogorzalek2019}, 
spin-squeezed states \cite{Leroux2010, Hosten-2016, Bao2020} 
and some variations of the Landau problem with a time-dependent magnetic field \cite{Dodonov2018,Dodonov2019}.
Moreover, when squeezed states of light are employed, an astonishing improvement in the detection rate at LIGO \cite{Barsotti2018, Tse2019} 
and a sensitivity enhancement in the shot noise limit at the Advanced Virgo gravitational wave detector \cite{Acernese2019} are observed.

The more general case of a driven TDHO has already been formally solved by different methods and approaches. 
The first solution came from Husimi in 1953 \cite{Husimi-1953} where he used an ansatz of a gaussian type 
to show that the formal solution of this problem can be obtained from the corresponding classical solution. 
Further important  contributions were made by Lewis and Riesenfeld \cite{LEWIS-1969}, Popov and Perelomov \cite{Popov-1969}, 
and Malkin,  Man'ko and collaborators \cite{Malkin-1970, Man'Ko-1970}. 
An extensive list of references in this line can be found in \cite{DODONOV-2005}. 
In these papers the authors introduced the use of invariants for time-dependent hamiltonians, 
a method still very used nowadays \cite{nagiyev-2019, zelaya-2020}. 
Algebraic solutions for the driven TDHO have been known since the 80's,
see for instance the papers of Ma and Rhodes \cite{Rhodes-1989}, with further contributions from Lo \cite{C.F.LO-1990}. 
In these papers it is shown that the time evolution operator (TEO) at any instant can be expressed as a product of 
a squeezing operator, a Glauber operator and a rotation operator, apart from an overall phase factor. 
Other authors have also investigated the TDHO considering different initial states and 
specific time-dependent parameters \cite{DODONOV-1979, ABDALLA-1985,CHENG-1988,GERRY-1990,CFLO-1990,CFLO-1991,Kumar-1991,
Twamley-1993,JANSZKY-1994,Pedrosa-1997,Pedrosa.I.A.-1997,Lima-2008,Buyukasik2019}. 
Some particular cases with exact known solutions, namely the sudden and linear frequency modulations have also been considered 
\cite{JANSZKY-1986, Kumar-1991, JANSZKY-1992, JANSZKY-TE-1994, MOYA-2003, 2019-AJP-Tiba}.

It is worth mentioning that previous solutions of the TDHO
based on algebraic methods are not very practical for numerical implementations, since the final expressions 
are written in terms of functions that satisfy non-linear differential equations involving the time-dependent 
parameters under consideration. In this work, our main purpose is to establish a procedure 
for solving this system in such a way that no matter which time-dependent functions for the parameters are considered, one will be able 
to calculate the quantum state of the system at any instant and with the desired precision for any initial state.
Using operator ordering techniques, similar to those used in Refs.\cite{truax-1985, CHENG-1988, Rhodes-1989, Rau-1996, paredes-2020}, 
based on BCH-like relations of the \textit{su}(1,1) Lie algebra and 
a time-splitting approach exploring the composition property of the TEO, 
we obtain an iterative analytical solution given by simple recurrence relations 
presented in the form of generalized continued fractions. The TEO is written in terms 
of the \textit{su}(1,1) Lie algebra generators and therefore is ready for application over 
any initial state. 
In order to prove the usefulness of our method we consider the HO initially in its fundamental state and 
we study its time-evolution for a variety of non-trivial frequency modulations. At first, these frequency modulations 
are chosen in order to reproduce some results from literature. 
Secondly, we analyse and compare two important cases: 
the parametric resonance modulation and the so-called Janszky-Adam scheme \cite{JANSZKY-1992}.

This paper is organized as follows. In Section \ref{Meth} 
we introduce the system studied and a detailed discussion of our method, the main theoretical result of this work. 
In Section \ref{RAD} we perform some numerical implementations and use our results to compare the efficiency on squeezing among different procedures. 
Section \ref{C} is left for the conclusions and final remarks.


\section{Time-dependent Harmonic oscillator}\label{Meth}

Let us consider a one-dimensional harmonic oscillator (HO) with arbitrary time-dependent parameters, namely, its mass $m(t)$ and frequency $\omega(t)$, whose hamiltonian is given by 
\begin{equation}
\hat{H}(t)=\frac{\hat{p}^{2}}{2m(t)}+\frac{1}{2}m(t)\omega^{2}(t)\hat{q}^2\, .
\label{eq:HOVF}
\end{equation}
Our purpose is to determine the time-evolution operator (TEO) of the HO at an arbitrary subsequent time.
However, this is not an easy task, since the hamiltonian is a time-dependent one and, hence, the computation of the TEO is quite intricate. 
In fact, from its definition
\begin{equation}
\left|\psi(t)\right\rangle = \hat{U}(t,0)\left|\psi(0)\right\rangle \, ,
\label{Definicao-U}
\end{equation}
and  the Schr\"odinger equation (we are using $\hbar = 1$), 
\begin{equation}
i\frac{\partial}{\partial t}\left|\psi(t)\right\rangle=\hat{H}(t)\left|\psi(t)\right\rangle\, ,
\label{eq:Schro-Eq}
\end{equation}
 it is immediate to  see that the TEO
satisfies the following differential equation,
\begin{equation}
i\frac{\partial}{\partial t}\hat{U}(t,0)=\hat{H}(t)\hat{U}(t,0)\, ,
\label{eq:Schro-TEO}
\end{equation}
 with the initial condition $ \hat{U}(0,0) = 1\!\! 1$, whose solution
can be written as the formal expression \cite{Sakurai-Book-2014}
\begin{equation}
\hat{U}(t,0) = T\left\{\exp\left[-i\int_0^t{\hat H}(t^\prime) dt^\prime\right]\right\}\, ,
\label{eq:DysonB}
\end{equation}
where $T$ means time ordering operator. This expression is known as Dyson series and 
its application to the problem at hand is extremely difficult. Instead of using Dyson series
it is more convenient, as we shall see, to appeal to the composition property of the TEO which follows directly from 
 definition (\ref{Definicao-U}), namely,  
\begin{equation}
\hat{U}(t,0)=\hat{U}(t,t_{N-1})\hat{U}(t_{N-1},t_{N-2})\cdots\hat{U}(t_{1},0)\, .
\label{eq:compoTEO}
\end{equation}
Although for finite time intervals the expressions $\hat{U}(t_j,t_{j-1})$, with $j = 1,2,...,N-1$, are quite involved since the problem under consideration has a time-dependent  hamiltonian (they are given, essentially by Dyson series), if we take 
$\tau\rightarrow 0$ ($\tau=t_{j}-t_{j-1}; j=1,2,...N$) and $N\rightarrow\infty$ with $N\tau = t$, we can write the TEO as an infinite product of simple infinitesimal time-evolution operators, namely, 
\begin{equation}
\hat{U}(t,0)  = \lim_{\substack{N\rightarrow\infty\\ N\tau = t}}e^{-i\hat{H}(N\tau)\tau}e^{-i\hat{H}\left((N-1)\tau\right)\tau}\cdots\;e^{-i\hat{H}(\tau)\tau} \, .
\label{eq:TEOin}
\end{equation}
Our iterative method is based on the above equation.
Let us introduce, as usual in algebraic methods for the HO,  the annihilation $\hat{a}$ and creation $\hat{a}^{\dagger}$ operators
\begin{equation}
\hat{a}\equiv\sqrt{\frac{m_{0}\omega_{0}}{2}}\left(\hat{q}+i\frac{\hat{p}}{m_{0}\omega_{0}}\right) \; ;\;\;\;\;
\hat{a}^{\dagger}\equiv\sqrt{\frac{m_{0}\omega_{0}}{2}}\left(\hat{q}-\frac{i}{m_{0}\omega_{0}}\hat{p}\right) \, ,
\label{eq:instantbasist0}
\end{equation}
where $m(t=0)\equiv m_{0}$, $\omega(t=0)\equiv\omega_{0}$ and $\left[\hat{a},\hat{a}^{\dagger}\right]=1$. 
We recall that we are working in the the Schr\"odinger picture. Therefore, inverting the above equations and substituting the expressions of operators $\hat p$ and $\hat q$ in terms of the operators $\hat{a}$ and $\hat{a}^{\dagger}$ into Eq. (\ref{eq:HOVF}), we obtain after a straightforward calculation
\begin{equation}
\hat{H}(t)=2\omega(t)\cosh\bigl[2\rho(t)\bigr]\hat{K}_{c}+\omega(t)\sinh\bigl[2\rho(t)\bigr]\left(\hat{K}_{+}+\hat{K}_{-}\right),\\
\label{eq:HamLie}
\end{equation}
where we defined
\begin{equation}
\rho(t)\equiv\frac{1}{2}\ln{\left(\frac{m(t)\omega(t)}{m_{0}\omega_{0}}\right)} \, .
\label{eq:compdgr}
\end{equation}
as well as the operators
\begin{equation}
\hat{K}_{+} := \frac{\hat{a}^{\dagger^{2}}}{2}, \:\: \hat{K}_{-} := \frac{\hat{a}^{2}}{2} \:\:\:\: \mbox{and} \:\:\:\: \hat{K}_{c} := 
\frac{\hat{a}^{\dagger}\hat{a}+\hat{a}\hat{a}^{\dagger}}{4}\, .\\
\label{eq:LieAlgebraGen}
\end{equation}
It is straightforward to show that the above operators satisfy the following commutation relations
\begin{equation}
\left[\hat{K}_{+},\hat{K}_{-}\right]=-2\hat{K}_{c} \:\: \mbox{and} \:\: \left[\hat{K}_{c},\hat{K}_{\pm}\right]=\pm\hat{K}_{\pm}\, ,
\label{eq:algebraK}
\end{equation}
so that they can be identified as the three generators of the \textit{su}(1,1) Lie algebra. This fact will allows us to use appropriate BCH-like formulas for this Lie algebra to obtain the TEO of the system. 
For future convenience, let us now introduce the so-called vacuum squeezed states (a detailed discussion can be found in Ref.\cite{Barnett-1997}). 
A single-mode vacuum squeezed state $\vert z\rangle$ (referred to, henceforth, simply by squeezed state) of the initial hamiltonian $\hat{H}_{0}\equiv\hat{H}(t=0)$, 
can be obtained by application of the squeezing operator ${\hat S}(z)$ on the fundamental state, $\vert z\rangle = {\hat S}(z) \vert 0\rangle$, with ${\hat S}(z)$ defined by
\begin{equation}
\label{gensqop}
\hat{S}(z)\equiv\exp \left\{-\frac{z}{2}\left.\hat{a}^{\dagger}\right.^{2}+\frac{z^{*}}{2}\hat{a}^{2}\right\} \, ,
\end{equation}
where $z = r e^{i\varphi}$ is a complex number.  With the aid of ordering theorems the squeezed state can be written as a superposition  
of the even energy eigenstates  \cite{Barnett-1997}
\begin{equation}
\left|z\right\rangle=\sqrt{\mbox{sech}(r)}\sum_{n=0}^{\infty}\frac{\sqrt{(2n)!}}{n!}
\left[-\frac{1}{2}e^{(i\varphi)}\tanh(r)\right]^{n}\left|2n\right\rangle.
\label{eq:squeezed}
\end{equation}
Note that $z$, and hence $r$ and $\varphi$, determines uniquely the squeezed state. In order to interpret $r$ and $\varphi$, it is convenient to introduce 
the quadrature operator ${\hat Q}_\lambda$, defined by  \cite{Barnett-1997}
\begin{equation}
\hat{Q}_{\lambda}=\frac{1}{\sqrt{2}}\left[e^{i\lambda} \hat{a}^{\dagger}+e^{-i\lambda} \hat{a}\right] \, .
\label{eq:quadratureop}
\end{equation}
The quadrature operators satisfy the commutation relation $[{\hat Q}_\lambda,{\hat Q}_{\lambda + \pi/2}] = i$. 
It is evident from the previous definition that 
$\hat{Q}_{\lambda=0} = (\hat{a}^{\dagger} +  \hat{a})/\sqrt{2} \propto   \hat{q}$ and 
$\hat{Q}_{\lambda=\pi/2} = i(\hat{a}^{\dagger} -  \hat{a})/\sqrt{2}  \propto   \hat{p}$. 
It can be shown that the variance of the quadrature operator in a squeezed state is given by \cite{Barnett-1997}
\begin{equation}
 \left(\Delta Q_\lambda \right)^2 = 
\frac{1}{2}\left[e^{2r} \sin^{2}\left(\lambda-\varphi/2\right)+e^{-2r} \cos^{2}\left(\lambda-\varphi/2\right)\right]\, 
\label{eq:quadratureopvariance}
\end{equation}
and the harmonic oscillator is said to be squeezed if the variance of one of the quadratures is smaller than $\frac{1}{2}$. Note the explicit dependence of $\left(\Delta Q_\lambda\right)^2$ with $r$ and $\varphi$. Further, from the previous equation we see that 
\begin{equation}
 \frac{e^{-2r}}{2} \le  \left(\Delta Q_\lambda \right)^2 \le  \frac{e^{2r}}{2}\, ,
\end{equation}
which justifies the interpretation of  $r$ as the squeezing parameter (SP). Parameter $\varphi$ is referred to as the squeezing phase (Sph).


\subsection{Time evolution}\label{Evl}

In this subsection we shall obtain the TEO of the system through an iterative method. To simplify calculations, but without any loss of generality, we consider the explicit time-dependence of the HO lying only in the frequency, such that $m(t)=1$. 

\subsubsection{Time-Splitting}\label{timespli}
Let us consider a time discretization in small intervals of equally size $\tau$ and let the frequency function $\omega(t)$ be considered constant in each of these intervals as follows \cite{JANSZKY-1994}
\begin{alignat}{2}
\omega(t)=\left\{
\begin{array}{ccc}
\omega_{0} & \mbox{for} & t\leq 0 \\
\omega_{1} & \mbox{for} & 0<t\leq\tau \\
 \vdots & & \vdots \\
\omega_{j} &\mbox{for} & (j-1)\tau<t\leq j\tau \\
 \vdots & & \vdots \\
\omega_{N} & \mbox{for} & (N-1)\tau<t\leq N\tau \\
\end{array} \right.
\label{eq:discrete}
\end{alignat}
where $\omega_j$ can be taken as any value assumed by $\omega(t)$ with $t_{j-1} < t \le t_j$. For convenience, we choose $\omega_{j} :=\omega(j\tau)$. 
Recall that $N\tau=t$ and an exact result is obtained only in the limit $N\rightarrow\infty$ ($\tau\rightarrow 0$). Note also that we chose time-dependent frequencies that are constant and equal to $\omega_0$ from $-\infty$ to $t=0$.
Once $t_j - t_{j-1} = \tau$, for any $j$, Eq. (\ref{eq:compoTEO}) takes the form
\begin{equation}
\hat{U}(t,0)=\hat{U}(N\tau,(N-1)\tau)\cdots\hat{U}(2\tau,\tau)\hat{U}(\tau,0) \, .
\label{eq:GenTEO}
\end{equation}
Assuming $N$ is as large as we want, we may approximate the hamiltonian  in each time interval $t_{j-1} < t \le t_j$, denoted by $H_j$,  as a constant one. Hence, from  
 Eq. (\ref{eq:HamLie}) we may write
\begin{equation}
\hat{H}_{j}=2\omega_{j}\cosh(2\rho_{j})\hat{K}_{c}+\omega_{j}\sinh(2\rho_{j})\left(\hat{K}_{+}+\hat{K}_{-}\right) \, ,
\label{eq:HamLiegen}
\end{equation}
where from Eq. (\ref{eq:compdgr}) it is clear that $\rho_{j}=\frac{1}{2}\ln{\left(\frac{\omega_{j}}{\omega_{0}}\right)}$.
Since all $H_j$ are now considered as time-independent hamiltonians, the TEO for each time interval, ${\hat U}(t_j,t_{j-1})$, with $j = 1,2,...,N$, can be written as $\hat{U}_{j} := {\hat U}(t_j,t_{j-1}) = e^{-i\hat{H}_{j}\tau}$. Therefore, using Eq. (\ref{eq:HamLiegen}), we can write
\begin{equation}
\hat{U}_{j}=e^{\lambda_{j+}\hat{K}_{+}+\lambda_{jc}\hat{K}_{c}+\lambda_{j-}\hat{K}_{-}},
\label{eq:TEOjk}
\end{equation}
where we defined
\begin{alignat}{1}
\label{truej1}
&\lambda_{j+} = \lambda_{j-}=-i\omega_{j}\tau \sinh(2\rho_{j}) \, , \\
\label{truej2}
&\lambda_{jc} = - 2i\omega_{j}\tau \cosh(2\rho_{j}) \, .
\end{alignat}
Using well known BCH relations of the $su(1,1)$ Lie algebras \cite{Barnett-1997, 2019-AJP-Tiba},  it is possible to write Eq. (\ref{eq:TEOjk}) as a product of exponentials of the Lie algebra generators in a suitable order, namely,
\begin{equation}
\hat{U}_{j}=e^{\Lambda_{j+}\hat{K}_{+}}e^{\ln(\Lambda_{jc})\hat{K}_{c}}e^{\Lambda_{j-}\hat{K}_{-}},
\label{eq:jTEO}
\end{equation}
where 
\begin{equation}
\Lambda_{jc}=\left(\cosh(\nu_{j})-\frac{\lambda_{jc}}{2\nu_{j}} \sinh(\nu_{j})\right)^{-2} \:\:\:\:\: \mbox{and} \:\:\:\:\:
\label{truej5}
\Lambda_{j\pm}=\frac{2\lambda_{j\pm} \sinh(\nu_{j})}{2\nu_{j} \cosh(\nu_{j})-\lambda_{jc}\sinh(\nu_{j})} \, ,
\end{equation}
with $\nu_{j}$ given by
\begin{equation}
\label{truej6}
\nu_{j}^{2} = \frac{1}{4}\lambda_{jc}^{2}-\lambda_{j+}\lambda_{j-}.
\end{equation}
Inserting Eqs. (\ref{truej1}) and (\ref{truej2}) into Eq. (\ref{truej6}) it is straightforward to show that $\nu_{j}=\pm i\omega_{j}\tau$, and substituting the obtained result into Eqs. (\ref{truej5}), consequently we obtain
\begin{alignat}{1}
\label{truej8}
& \Lambda_{jc}=\left(\cos(\omega_{j}\tau)+i\cosh(2\rho_{j})\sin(\omega_{j}\tau)\right)^{-2}, \\
\label{truej9}
&\Lambda_{j\pm}=\frac{-i \sinh(2\rho_{j})\sin(\omega_{j}\tau)}{\cos(\omega_{j}\tau)+i\cosh(2\rho_{j})\sin(\omega_{j}\tau)}.
\end{alignat}
Therefore, using Eqs. (\ref{Definicao-U}), (\ref{eq:GenTEO}) and (\ref{eq:jTEO}) the state of the system at an arbitrary instant $t>0$ 
can be written as the following product of operators
\begin{alignat}{1}
\left|\psi(t)\right\rangle =& e^{\Lambda_{N+}\hat{K}_{+}}e^{\ln(\Lambda_{Nc})\hat{K}_{c}}e^{\Lambda_{N-}\hat{K}_{-}} \, e^{\Lambda_{(N-1)+}\hat{K}_{+}}e^{\ln(\Lambda_{(N-1)c})\hat{K}_{c}}e^{\Lambda_{(N-1)-}\hat{K}_{-}}\;  \cdots\cr
&\cdots\;  e^{\Lambda_{2+}\hat{K}_{+}}e^{\ln(\Lambda_{2c})\hat{K}_{c}}e^{\Lambda_{2-}\hat{K}_{-}} \, e^{\Lambda_{1+}\hat{K}_{+}}e^{\ln(\Lambda_{1c})\hat{K}_{c}}e^{\Lambda_{1-}\hat{K}_{-}}\left|\psi(0)\right\rangle.
\label{eq:finalstate}
\end{alignat}
This formula is not yet suitable for numerical applications but, as we shall see in the next subsection, it is the starting point for the deduction of a very convenient recurrence relation which will be the core of our iterative method.


\subsubsection{Iterative method and TEO}\label{RF}

Here we shall show that the composition of $N$ operators $U_{j}$ of the form of Eq. (\ref{eq:jTEO}), 
can be written as another operator with the same form, as is the case of the TEO in Eq. (\ref{eq:finalstate}).
In fact, this is possible since any $U_{j}$ is an element of the Lie group and, as it is well known, 
the composition of two or more elements of a group yields to another element of the group \cite{GILMORE-2012}.
Since we assume $\omega(t)$ is constant by parts, as defined in Eq. (\ref{eq:discrete}),
then any change in the frequency is abrupt, \textit{i.e}, a jump. 
It is worth mentioning that a HO with a jump in the frequency is one of the few cases 
of frequency modulation with analytical solution. 
A detailed study of the later can be found in Ref.\cite{2019-AJP-Tiba}.

In order to obtain a recurrence formula we initially calculate the composition of the first two elements 
present in the TEO of Eq. (\ref{eq:finalstate}), namely,
\begin{equation}
\hat{U}(2\tau,0)=e^{\Lambda_{2+}\hat{K}_{+}}e^{\ln(\Lambda_{2c})\hat{K}_{c}}e^{\Lambda_{2-}\hat{K}_{-}}\, e^{\Lambda_{1+}\hat{K}_{+}}e^{\ln(\Lambda_{1c})\hat{K}_{c}}e^{\Lambda_{1-}\hat{K}_{-}} \, .
\label{eq:GenTEO2}
\end{equation}
Since we want to write the above operator product as another $U_{j}$, 
we shall use ordering techniques to move the operators until obtaining the desired configuration.
In the following discussion we will use the well known BCH relations \cite{Barnett-1997}
\begin{equation}
\label{eq:BCHclasic}
e^{\hat{A}}\hat{B}e^{-\hat{A}}=\hat{B}+\left[\hat{A},\hat{B}\right]+\frac{1}{2!}\left[\hat{A},\left[\hat{A},\hat{B}\right]\right]+
\frac{1}{3!}\left[\hat{A},\left[\hat{A},\left[\hat{A},\hat{B}\right]\right]\right]+\ldots \, ,
\end{equation}
and
\begin{equation}
e^{\hat{A}}f\left(\hat{C}\right)e^{-\hat{A}}=f\left(e^{\hat{A}}\hat{C}e^{\hat{-A}}\right) \, .
\label{eq:BCHfunc}
\end{equation}
Using the identity as $\mathds{1}=e^{\Lambda_{1+}\hat{K}_{+}}e^{-\Lambda_{1+}\hat{K}_{+}}$ and Eq. (\ref{eq:BCHfunc}) in Eq. (\ref{eq:GenTEO2}), we obtain
\begin{alignat}{1}
\hat{U}(2\tau,0)=&e^{\Lambda_{2+}\hat{K}_{+}}e^{\ln(\Lambda_{2c})\hat{K}_{c}} e^{\Lambda_{1+}\hat{K}_{+}}\, \left(e^{-\Lambda_{1+}\hat{K}_{+}}e^{\Lambda_{2-}\hat{K}_{-}}e^{\Lambda_{1+}\hat{K}_{+}}\right) \, e^{\ln(\Lambda_{1c})\hat{K}_{c}}e^{\Lambda_{1-}\hat{K}_{-}}  \cr
=&e^{\Lambda_{2+}\hat{K}_{+}}e^{\ln(\Lambda_{2c})\hat{K}_{c}} e^{\Lambda_{1+}\hat{K}_{+}}\, \left(e^{\left\{\sigma_{+}\hat{K}_{+}+\sigma_{c}\hat{K}_{c}+\sigma_{-}\hat{K}_{-}\right\}}\right) \, e^{\ln(\Lambda_{1c})\hat{K}_{c}}e^{\Lambda_{1-}\hat{K}_{-}} \, ,
\label{eq:GenTEO3}
\end{alignat}
where 
\begin{equation}
\sigma_{+}=\Lambda_{2-}(\Lambda_{1+})^{2}  \:\:\: \mbox{;} \:\:\: \sigma_{c}=2\Lambda_{2-}\Lambda_{1+} \:\:\: \mbox{and} \:\:\: 
\sigma_{-}=\Lambda_{2-} \, .
\label{eq:sigmas}
\end{equation}
Now, for the embraced exponential in Eq. (\ref{eq:GenTEO3}) we can use the factorised representation Eq. (\ref{eq:jTEO}) to write
\begin{alignat}{1}
\hat{U}(2\tau,0)=&e^{\Lambda_{2+}\hat{K}_{+}}e^{\ln(\Lambda_{2c})\hat{K}_{c}} e^{\Lambda_{1+}\hat{K}_{+}}\, \left(e^{\Sigma_{+}\hat{K}_{+}}e^{\ln(\Sigma_{c})\hat{K}_{c}}e^{\Sigma_{-}\hat{K}_{-}}\right) \, e^{\ln(\Lambda_{1c})\hat{K}_{c}}e^{\Lambda_{1-}\hat{K}_{-}}  \cr
=&e^{\Lambda_{2+}\hat{K}_{+}}\,\left(e^{\ln(\Lambda_{2c})\hat{K}_{c}} e^{(\Lambda_{1+}+\Sigma_{+})\hat{K}_{+}}\right)\, e^{\ln(\Sigma_{c})\hat{K}_{c}}e^{\Sigma_{-}\hat{K}_{-}} e^{\ln(\Lambda_{1c})\hat{K}_{c}}e^{\Lambda_{1-}\hat{K}_{-}} \, ,
\label{eq:GenTEO4}
\end{alignat}
where 
\begin{equation}
\label{eq:Sigmapfin}
\Sigma_{+}=\frac{(\Lambda_{1+})^{2}\Lambda_{2-}}{1-\Lambda_{2-}\Lambda_{1+}} \, , \:\:\:\:
\Sigma_{c}=\left(1-\Lambda_{2-}\Lambda_{1+}\right)^{-2} \:\:\:\:  \mbox{and} \:\:\:\: 
\Sigma_{-}=\frac{\Lambda_{2-}}{1-\Lambda_{2-}\Lambda_{1+}} \, .
\end{equation}
Following a similar protocol, \textit{i.e.}, insertion of the identity as
$\mathds{1}=e^{-\ln(\Lambda_{2c})\hat{K}_{c}}e^{\ln(\Lambda_{2c})\hat{K}_{c}}$ 
followed by the use of Eqs. (\ref{eq:BCHclasic}) and (\ref{eq:BCHfunc}), 
the embraced quantity in Eq. (\ref{eq:GenTEO4}) results in
\begin{alignat}{1}
\hat{U}(2\tau,0)=& e^{\Lambda_{2+}\hat{K}_{+}}\,\left(e^{\left\{(\Lambda_{1+}+\Sigma_{+})\Lambda_{2c}\,\hat{K}_{+}\right\}}e^{\ln(\Lambda_{2c})\hat{K}_{c}}\right)\, e^{\ln(\Sigma_{c})\hat{K}_{c}}e^{\Sigma_{-}\hat{K}_{-}} e^{\ln(\Lambda_{1c})\hat{K}_{c}}e^{\Lambda_{1-}\hat{K}_{-}}  \cr
=&e^{\left\{\Lambda_{2+}+\Lambda_{2c}(\Lambda_{1+}+\Sigma_{+})\right\}\hat{K}_{+}}e^{\ln(\Lambda_{2c}\Sigma_{c})\hat{K}_{c}}
 \left(e^{\Sigma_{-}\hat{K}_{-}}e^{\ln(\Lambda_{1c})\hat{K}_{c}}\right)e^{\Lambda_{1-}\hat{K}_{-}}\, .
\label{eq:GenTEO5}
\end{alignat}
As before, the embraced exponential product in the above equation is reordered by introducing 
this time the identity as $\mathds{1}=e^{\ln(\Lambda_{1c})\hat{K}_{c}}e^{-\ln(\Lambda_{1c})\hat{K}_{c}}$ 
followed by the use of Eqs. (\ref{eq:BCHclasic}) and (\ref{eq:BCHfunc}) as
\begin{alignat}{1}
\hat{U}(2\tau,0)=& e^{\left\{\Lambda_{2+}+\Lambda_{2c}(\Lambda_{1+}+\Sigma_{+})\right\}\hat{K}_{+}}e^{\ln(\Lambda_{2c}\Sigma_{c})\hat{K}_{c}}
 \left(e^{\ln(\Lambda_{1c})\hat{K}_{c}}e^{\Sigma_{-}\Lambda_{1c}\hat{K}_{-}}\right)e^{\Lambda_{1-}\hat{K}_{-}}\, \cr
=&e^{\bigl(\Lambda_{2+}+\Lambda_{2c}(\Lambda_{1+}+\Sigma_{+})\bigl)\hat{K}_{+}}e^{\ln(\Lambda_{2c}\Lambda_{1c}\Sigma_{c})\hat{K}_{c}} 
e^{(\Sigma_{-}\Lambda_{1c}+\Lambda_{1-})\hat{K}_{-}}\, .
\label{eq:GenTEO6}
\end{alignat}
Finally, substitution of the big sigmas, Eq. (\ref{eq:Sigmapfin}), in the above equation results in
\begin{equation}
\hat{U}(2\tau,0)=
e^{(\Lambda_{2+}+\frac{\Lambda_{2c}\Lambda_{1+}}{1-\Lambda_{2-}\Lambda_{1+}})\hat{K}_{+}}
e^{\ln\left(\frac{\Lambda_{2c}\Lambda_{1c}}{(1-\Lambda_{2-}\Lambda_{1+})^{2}}\right)\hat{K}_{c}}
e^{(\Lambda_{1-}+\frac{\Lambda_{2-}\Lambda_{1c}}{1-\Lambda_{2-}\Lambda_{1+}})\hat{K}_{-}} \, .
\label{eq:2compfin}
\end{equation}
The recurrence relation for the composition of $N$ operators $U_{j}$ 
is a natural consequence of the previous result. 
For convenience, let us first define
\begin{equation}
\alpha_{1}=\Lambda_{1+} \:\:\:\:\: \mbox{;} \:\:\:\:\: \beta_{1}=\Lambda_{1c}  \:\:\:\:\: \mbox{and} \:\:\:\:\: \gamma_{1}=\Lambda_{1-} \, .
\label{eq:alphagammabeta1}
\end{equation}
Then, if we write Eq. (\ref{eq:2compfin}) as
\begin{equation}
\hat{U}(2\tau,0)=e^{\alpha_{2}\hat{K}_{+}}e^{\ln(\beta_{2})\hat{K}_{c}}e^{\gamma_{2}\hat{K}_{-}} \, , \\
\label{eq:2compgamalta}
\end{equation}
comparison of the last two equations with Eq. (\ref{eq:2compfin}) allows us to make the identifications
\begin{equation}
\alpha_{2}=\Lambda_{2+}+\frac{\alpha_{1}\Lambda_{2c}}{1-\alpha_{1}\Lambda_{2-}} \:\:\: \mbox{,} \:\:\:\:\:\:
\beta_{2}=\frac{\beta_{1}\Lambda_{2c}}{\left(1-\alpha_{1}\Lambda_{2-}\right)^{2}} \:\:\:\:\: \mbox{and} \:\:\:\:\:
\label{beta2}
\gamma_{2}=\gamma_{1}+\frac{\Lambda_{2-}\beta_{1}}{1-\alpha_{1}\Lambda_{2-}} \, .
\end{equation}
It is worth emphasizing that the above relations are obtained no matter the ordering followed to move the operators, 
as long as the final operator have the structure of an $U_{j}$. Therefore, from Eqs. (\ref{eq:2compgamalta}) and (\ref{beta2}), the TEO is given by
\begin{equation}
\hat{U}(t,0)=
e^{\alpha_{N}\hat{K}_{+}}e^{\ln(\beta_{N})\hat{K}_{c}}e^{\gamma_{N}\hat{K}_{-}} \, ,
\label{eq:GenTEO1}
\end{equation}
where coefficients $\alpha$, $\beta$ and $\gamma$ are given by the following expressions
\begin{equation}
\alpha_{N}=\Lambda_{N+}+\frac{\alpha_{(N-1)}\Lambda_{Nc}}{1-\alpha_{(N-1)}\Lambda_{N-}} \:\:\: \mbox{,} \:\:\:\:\:\:
\beta_{N}=\frac{\beta_{(N-1)}\Lambda_{Nc}}{\left(1-\alpha_{(N-1)}\Lambda_{N-}\right)^{2}} \:\:\:\:\: \mbox{and} \:\:\:\:\:
\label{eq:betaN}
\gamma_{N}=\gamma_{(N-1)}+\frac{\Lambda_{N-}\beta_{(N-1)}}{1-\alpha_{(N-1)}\Lambda_{N-}} \, .
\end{equation}
We recall that $N\tau=t$ and the exact result is obtained by taking the limit $N\rightarrow\infty$ ($\tau\rightarrow 0$) in the above expressions. 
However, since to perform numerical calculations we must set the value of $N$, once its value is found that guarantees the convergence, we can say   
that our result is as exact as our computer allows us to increase $N$ above it convergence value. From Eqs. (\ref{eq:finalstate}) and (\ref{eq:GenTEO1}) we obtain the state vector
\begin{equation}
\left|\psi(t)\right\rangle =e^{\alpha_{N}\hat{K}_{+}}e^{\ln(\beta_{N})\hat{K}_{c}}e^{\gamma_{N}\hat{K}_{-}} \left|\psi(0)\right\rangle.
\label{eq:finalstatef}
\end{equation}
Note that the coefficient $\alpha$ in Eq. (\ref{eq:GenTEO1}) is independent of $\beta$ and $\gamma$. 
This enables us to write it in the following convenient form
\begin{equation}
\alpha_{j}=\Lambda_{j+}-\frac{\Lambda_{jc}}{\Lambda_{j-}-\frac{1}{\Lambda_{(j-1)+}-
\frac{\Lambda_{(j-1)c}}
{\Lambda_{(j-1)-} \, -\frac{1}{\ddots \Lambda_{2+}-\frac{\Lambda_{2c}}{\Lambda_{2-}-\frac{1}{\Lambda_{1+}}}}}}} \\
\label{eq:gammarecursive}
\end{equation}
The above expression is a generalized continued fraction (GCF) for which some topics such as convergence can be investigated. 
Notice that Eq. (\ref{eq:gammarecursive}) enables an easy numerical implementation. GCFs lie in the context of complex analysis and are specially useful to study analyticity of functions as well as number theory among other fields. For an interested reader we suggest Ref.\cite{Kinchin-1997}.

Note that, from Eq. (\ref{eq:finalstatef}) we are able to calculate the final state provided any initial state is given. 
However, to do numerical calculations and in order to prove the usefulness of our method, we shall consider 
the initial state as the fundamental state $\left|\psi(0)\right\rangle=\left|0\right\rangle$. 
This will enables us to reproduce and analyse some well known results from literature.

From Eq. (\ref{eq:LieAlgebraGen}) we have the following results
\begin{equation}
\hat{K}_{-}\left|n\right\rangle=\frac{1}{2}\sqrt{n(n-1)}\left|n-2\right\rangle \:\:\: \mbox{,} \:\:\:\:\:\:
\hat{K}_{+}\left|n\right\rangle=\frac{1}{2}\sqrt{(n+1)(n+2)}\left|n+2\right\rangle \:\:\:\:\: \mbox{and} \:\:\:\:\:
\hat{K}_{c}\left|n\right\rangle=\frac{1}{2}\left(n+\frac{1}{2}\right)\left|n\right\rangle \, .
\label{eq:rules}
\end{equation}
Using the above equations and the well known expansion $e^{\hat{A}}=\sum_{n=0}^{\infty}{\frac{1}{n!}\hat{A}^{n}}$ 
(valid for a general operator $\hat{A}$) in Eq. (\ref{eq:finalstatef}), it can be shown that
\begin{equation}
\left|\psi(t)\right\rangle=\sqrt{\left|(\beta_{N})\right|^{1/2}}\sum_{n=0}^{\infty}\frac{\sqrt{(2n)!}}{n!}
\left[\frac{1}{2}\left|\alpha_{N}\right|e^{i\vartheta_{N}}\right]^{n}\left|2n\right\rangle,
\label{eq:wfinalg2}
\end{equation}
where the overall phase was removed by the redefinitions
\begin{equation}
\alpha_{N}=\left|\alpha_{N}\right|e^{i\vartheta_{N}}, \:\:\:\: \mbox{and}\:\:\:\: \beta_{N}=\left|\beta_{N}\right|e^{i\chi_{j}}.
\label{eq:phasecoeffg2}
\end{equation}
Note that, for any $j=1,2,...,N$, the relation
\begin{equation}
\left|\alpha_{j}\right|^{2}+\left|\beta_{j}\right| = 1
\label{eq:modulustrig}
\end{equation}
must be satisfied, as it can be straightforwardly checked.
Summing up: we have shown that any change in the frequency (as well as in the mass) of a HO, initially in its fundamental state, 
brings the system into a squeezed state of the initial hamiltonian $\hat{H}_{0}$: $\left|\psi(t)\right\rangle=\left|z(t)\right\rangle$, with $z(t)=r(t) e^{i\varphi(t)}$. The corresponding SP and Sph can be computed by comparing  Eqs. (\ref{eq:wfinalg2}) and (\ref{eq:squeezed}), a procedure which leads to
\begin{equation}
r(t)=\tanh^{-1}\left|\alpha_{N}\right|, \:\:\:\: \mbox{and}\:\:\:\: \varphi(t)=\vartheta_{N}\pm n\pi \:\: \mbox{with} \:\:(n=1,2,...)
\label{eq:phaseparamsque}
\end{equation}
Observe that the state is totally defined by the complex coefficient $\alpha_{j}$, since its modulus gives the SP 
$r(t)$ and its phase $\vartheta_N$ gives the Sph $\varphi(t)$.


\section{Numerical implementations and discussions} \label{RAD}

In this section, we will apply our iterative method, described previously, for a variety of frequency modulations. 
The numerical calculations will be implemented in the platform \textit{Mathematica}, 
where each frequency function will be discretized using very small intervals. We have established the optimal amount of such intervals for a given frequency function by studying the convergence of the method, so we can assure that the minimum used of 150.000 points (jumps) is good enough to perform a physical analysis.

Initially, in order to check the consistency of our method, 
in subsection \ref{AR} we recover some interesting results for the variances obtained by Adams and Janszky 
in Ref.\cite{JANSZKY-1994} by using a non-trivial frequency modulation. 
In this reference, the authors  considered time-dependent frequencies that 
return asymptotically to their original values as $t\rightarrow \infty$. 
In the last subsection, we consider time-dependent frequencies corresponding to 
two very efficient ways of generating squeezing states, namely, 
the parametric resonance modulation and the so-called Janszky-Adam (J-A) scheme \cite{JANSZKY-1992}.  
Choosing appropriately the parameters for both frequency modulations, we compare them and show that 
squeezing with the Adams-Janszky scheme is more efficient. 

In the following discussion, we shall use the particular definition of the scaled quadrature operators 
used by Janszky in Ref.\cite{JANSZKY-1994}, given by $1/\sqrt{2}$ times the quadrature operator 
defined in Eq. (\ref{eq:quadratureop}). As a consequence, 
squeezing occurs when the variance is smaller than the coherent limit given by $1/4$ instead of $1/2$.
Despite our results are valid for any initial frequency, 
for convenience we choose in all our numerical calculations $\omega_{0}=1$. 
Also we will use dimensionless quantities in the analysis.

\subsection{Checking the method}\label{AR}

\begin{figure}
\centering
\includegraphics*[width=16.0cm]{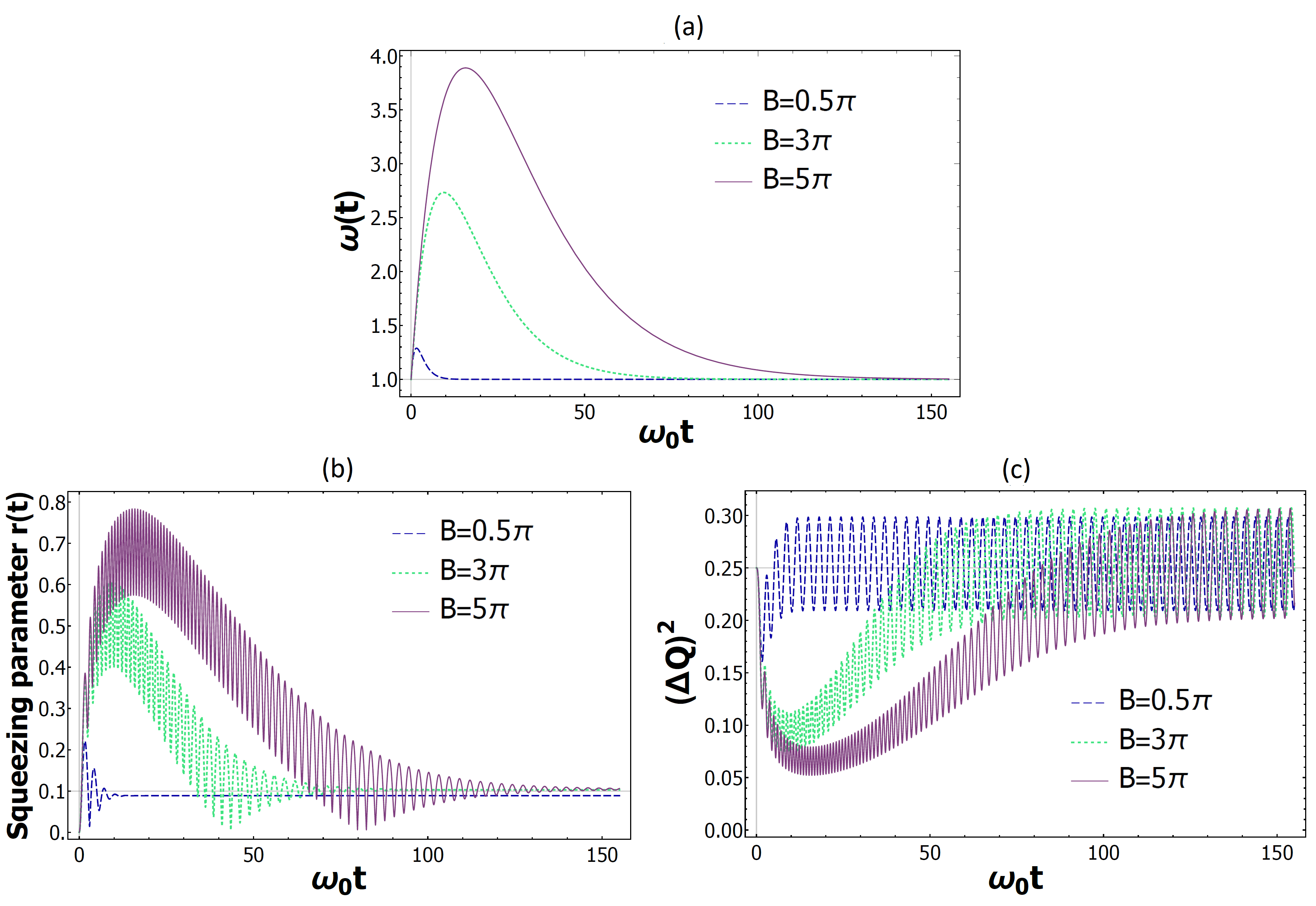}
\caption{From Eq. (\ref{eq:frequency2}) are plotted in the same time interval: (a) the frequency functions and, the time-evolution of (b) the SP and of (c) the quadrature variance. The curves associated to the three different values of the parameter are $B=0.5\pi$ (dashed line), $B=3\pi$ (dotted line) and $B=5\pi$ (solid line).}
\label{fig:ExponentialDecres}
\end{figure}

In order to get confidence in our method, in this subsection we will recover a well known result of the literature involving a harmonic oscillator with a time-dependent frequency. 
Our main purpose here is to obtain the squeezing parameter as well as the variance of a quadrature operator  for the system discussed in Ref. \cite{JANSZKY-1994}. Following this reference we consider a non-oscillatory frequency function of the type:
\begin{equation}
\omega(t)=\left\{
\begin{array}{ll}
\:\:\:\:\:\:\:\:\:\:\:\:\:\:\:\:\:\:\:\: \omega_{0} \:\:\:\:\:\:\:\:\:\:\:\:\:\:\:\:\:\:\:\:\:\:\:\:\:\:\:\: \mbox{for} \:\:\:\:\:\: t\leq 0 \cr
\omega_{0}\left[1+ \frac{\omega_{0} t}{2} \exp\left(-\frac{\omega_{0} t}{B}\right)\right] \:\:\:\:\:\:\:\:\:\:\: \mbox{for} \:\:\:\:\:\: t>0\, ,
\end{array} \right.
\label{eq:frequency2}
\end{equation}
where $B$ is a dimensionless positive parameter. 
In Fig.(\ref{fig:ExponentialDecres}.a)  we plot the previous frequency as a function of time for different values of $B$. Note that these frequencies are 
functions that start increasing, but after passing by their maximum values, they approach monotonically and asymptotically their original values. Also, note that the larger the parameter $B$, the longer it takes the frequency to return to its initial value $\omega_0$. In Fig.(\ref{fig:ExponentialDecres}.b), applying the method developed previously, we plot the squeezing parameter $r(t)$ as a function of time for the three frequencies plotted in Fig.(\ref{fig:ExponentialDecres}.a). We see that $r(t)$ has an oscillatory behavior which crudely follows the shape of the corresponding time-dependent frequency. Notice that the oscillations in $r(t)$ tend to cease as the frequency asymptotically returns to its  original value and the squeezing parameter evolves to a constant value which depends on $B$. This is a direct consequence of the fact that the final value of the frequency is the same as the initial one. 
In Fig.(\ref{fig:ExponentialDecres}.c) applying again our method we plot the time-evolution of the quadrature variance. This variance has an oscillatory behavior even after the  oscillations in the squeezing parameter tends to cease. This oscillatory behavior even after $r(t)$ has achieved a constant value is due solely to the squeezing phase term 
present in Eq. (\ref{eq:quadratureopvariance}). 

Our results are in total agreement with those appearing in Janszky's paper  \cite{JANSZKY-1994}. 
In fact, we have extended our analysis into a larger time-interval, from $\omega_{0}t=30$ (Janszky's paper) to $\omega_{0}t=150$, 
enabling us to see the behaviour in the asymptotic limit.
Since the time-dependent frequencies considered in this subsection are quite non-trivial, we can be very confident with our method and results. 
It is worth mentioning that in Janszky's paper  \cite{JANSZKY-1994} only the time-dependence of the quadrature is plotted, but not the time-dependence of the SP. 


\subsection{Parametric Resonance}\label{PR}

\begin{figure}
\centering
\includegraphics*[width=16.0cm]{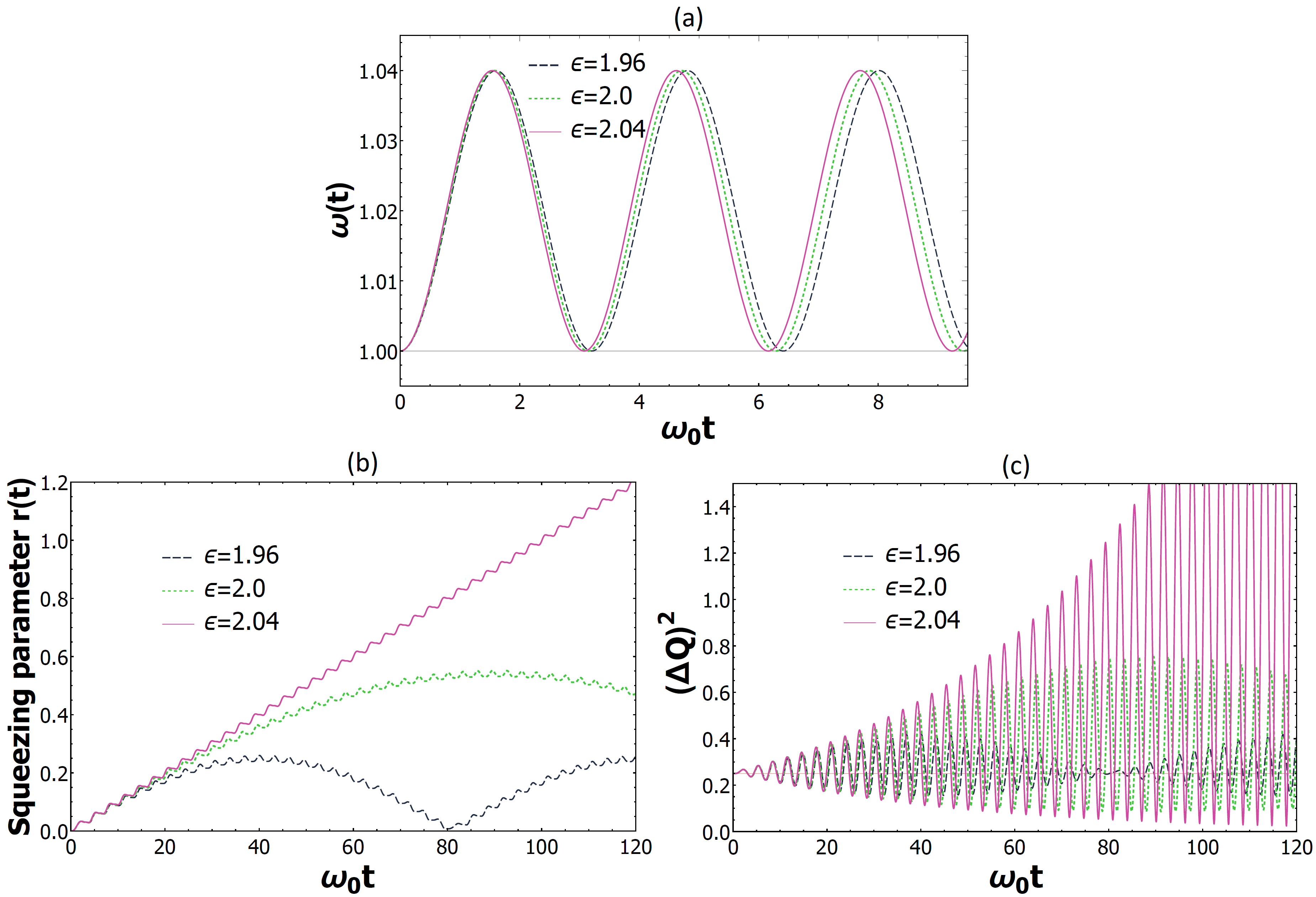}
\caption{(a) The frequency functions and, in the same time interval, the time-evolution of (b) the SP and (c) the quadrature variance. In (b), it is also shown the behavior of SP for the initial dynamics. The curves associated to the three different values of the parameter are $\epsilon=1.96$ (dashed line), $\epsilon=2.0$ (dotted line) and $\epsilon=2.04$ (solid line and resonance case).}
\label{fig:ParRes}
\end{figure}

\begin{figure}
\centering
\includegraphics*[width=16cm]{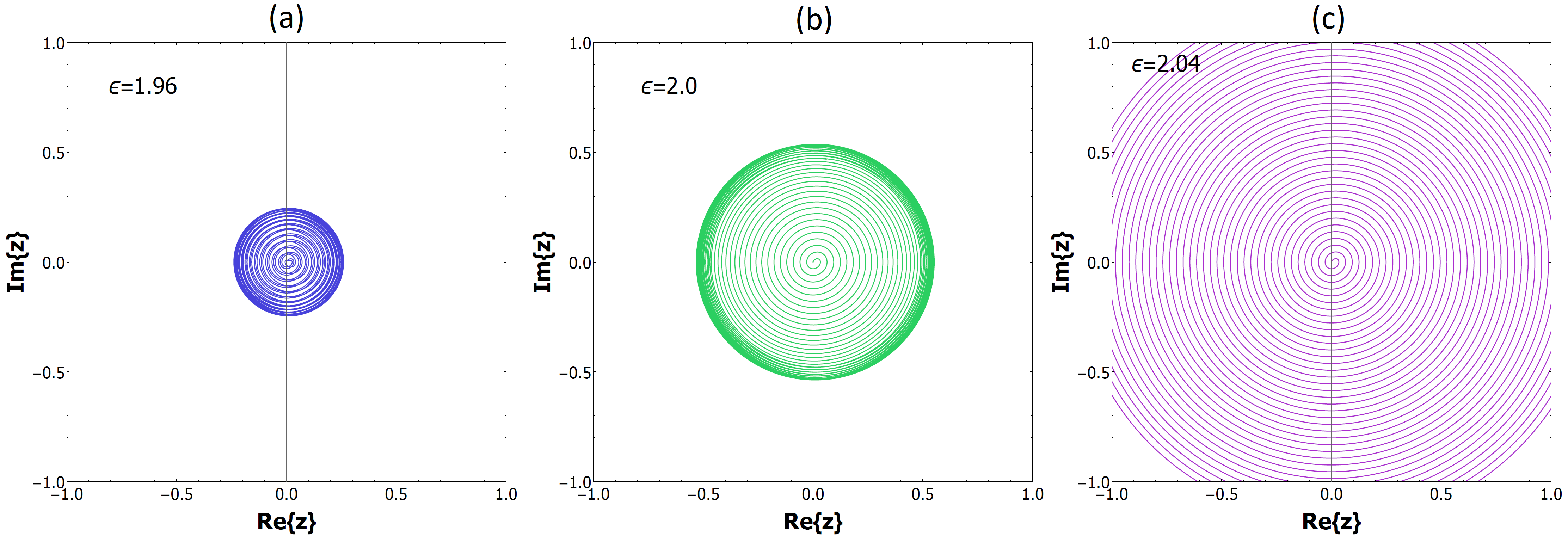}
\caption{Plot of the time-evolution in the interval $0\leq t\leq 120$ of the complex number $z$ characterizing the final state of the system for the frequency function given in Eq. (\ref{eq:frequency3}) and for the different values of the parameter: (a) $\epsilon=1.96$, (b) $\epsilon=2.0$ and (c) $\epsilon=2.04$ (resonance case).}
\label{fig:Resonance}
\end{figure}

Here we shall use a pulsating frequency function in the study of quantum parametric resonance. We shall show that in the resonance condition the mean value of the SP grows linearly with time, showing a certain characteristic angular coefficient, while for the non-resonance cases it has oscillatory behaviour (like beatings), with larger amplitude and period when closer to the resonance condition. After that, we also characterize the complex map (fingerprint) of the final state. 
Let us consider the following frequency function
\begin{equation}
\omega(t)=\left\{
\begin{array}{ll}
\:\:\:\:\:\:\:\:\:\:\:\:\:\:\:\:\:\:\:\:\:\:\:\:\:\:\:\:\:\:\:\:\:  \omega_{0} \:\:\:\:\:\:\:\:\:\:\:\:\:\:\:\:\:\:\:\:\:\:\:\:\:\:\:\:\:\:\:\:\:\:\:\:\:\: \mbox{for} \:\:\:\:\:\: t\leq 0 \cr\cr
\frac{1}{2}\left[\left(\omega_{0}+\omega_{l}\right)+\left(\omega_{0}-\omega_{l}\right) 
\cos\left(\epsilon \omega_{0}t\right)\right] \:\:\:\:\:\:\:\:\:\:\: \mbox{for} \:\:\:\:\:\:  t > 0
\end{array} \right.
\label{eq:frequency3}
\end{equation}
where $\omega_{l}$ is the maximum value reached by $\omega(t)$. Recall that $\omega_{0}=1$ and we fixed $\omega_{l}=1.04$, both in arbitrary units, for latter convenience. In Fig.(\ref{fig:ParRes}.a) we plot the above time-dependent frequency as a function of $\omega_{0}t$ for different values of the dimensionless parameter $\epsilon$. This parameter allows one to tune the parametric resonance phenomenon which occurs when $\epsilon\omega_{0}$ equals 
twice the value denoted by reference frequency $\omega_R = \frac{\omega_{0}+\omega_{l}}{2}$, 
which is the time average value of the harmonic oscillator frequency $\omega(t)$. 
The parametric resonance condition is then achieved with $\epsilon=2.04$. 
The other values of $\epsilon$ were chosen so that they are close but smaller than the resonant value\footnote{We could have chosen values for $\epsilon$ close but greater than $2.04$, but the results would have been the same.}.
In Fig.(\ref{fig:ParRes}.b) we plot the SP as a function of time for different values of $\epsilon$. The main characteristic shown in this figure is that at the resonance condition, the average value of  the SP grows linearly with time, indefinitely, in contrast to what happens in the non-resonance cases, where the average value of the SP starts growing, achieves a maximum value and then diminishes until it vanishes and then starts the process again, presenting a periodic behaviour.  Note that as $\epsilon$ approaches the resonant value, $\epsilon = 2.04$, the period of oscillation of $r(t)$ becomes larger, tending to infinity as $\epsilon \rightarrow 2.04$. It is worth mentioning that at the resonance condition the average energy is always increasing. 
In Fig.(\ref{fig:ParRes}.c) we plot the variance of the chosen quadrature as a function of time. Note that the characteristic  time of the variance oscillations is associated with the squeezing phase dynamics, while the amplitudes of those oscillations are related to the mean value of the SP (see Eq. (\ref{eq:quadratureopvariance})). As a consequence, the non-resonant cases exhibit  modulations as beats while at parametric resonance, since the average SP grows linearly with time, the modulations of the oscillations (the envelopes enclosing the oscillations) are exponentially increasing at the top of the oscillations and exponentially decreasing at the bottom of the oscillations. 
In Fig.(\ref{fig:Resonance}) we used Eq. (\ref{eq:phaseparamsque}) to plot, in the complex plane, the dynamics of the complex function $z = r e^{i\varphi} = \tanh^{-1}\left|\gamma\right| e^{i(\vartheta\pm\pi)}$, where $\left|z\right\rangle=\left|r e^{i \phi}\right\rangle$ is the (squeezed) state of the system, as it was similarly done  in Ref.\cite{GERRY-1990}. This is a geometric representation containing all relevant information about the dynamics of the system and can be considered as a fingerprint of the dynamics of the state. The non-resonant cases are plotted in Figs.(\ref{fig:Resonance}.a) and (\ref{fig:Resonance}.b). For these cases, the curves are limited  and bounded by a maximum radii in the complex plane of $z$, since, when out of resonance, the squeezing parameter ($r = \vert z\vert$) has a maximum value. The closer to the resonant condition, the larger the radius in the complex plane of the curves  that describes the dynamics of the system. This can be seen by inspection of  Figs.(\ref{fig:Resonance}.a) and (\ref{fig:Resonance}.b), since the larger radius corresponds to the larger value of $\epsilon$. At the resonance condition, as shown in Fig.(\ref{fig:Resonance}.c),  there is no enclosing circle and the curve is given by a growing spiral since $|z|$ increases exponentially indefinitely. 



\subsection{Janszky-Adam scheme $\times$ parametric resonance}\label{JAS}

\begin{figure}
\centering
\includegraphics*[width=15.0cm]{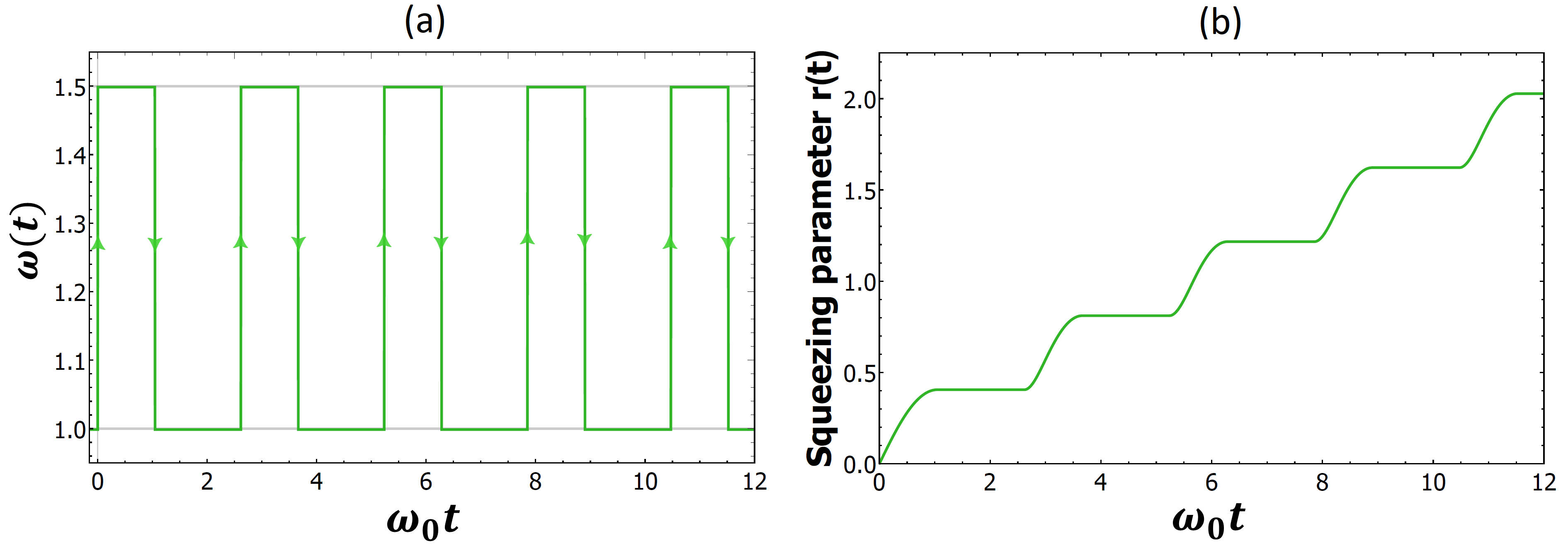}
\caption{(a) Plot of the frequency modulation function of the J-A scheme. (b) Time-evolution of the SP.}
\label{fig:JanzskyModel}
\end{figure}

The Janszky-Adam (J-A) scheme is known as a very strong squeezing model by frequency modulation in the harmonic oscillator \cite{FUJII-2015} and useful, for instance, in the description of a confined light field strongly coupled to a two-level system, or qubit, in the dispersive regime \cite{joshi-2017}. It uses sudden jumps between two fixed frequencies appropriately synchronized \cite{JANSZKY-TE-1994, JANSZKY-1992} and these abrupt frequency changes produce a high degree of squeezing  \cite{Kumar-1991}.
In Fig.(\ref{fig:JanzskyModel}.a), the time-dependent frequency of the HO in the Janszky-Adam model is plotted. It consists, as mentioned above, of  periodic sudden jumps between two constant frequencies, namely $\omega_{0}$ and $\omega_{1}$ (chosen to be 1.0 and 1.5 in arbitrary units in Fig.(\ref{fig:JanzskyModel}.a)). The respective time intervals in each frequency are suitable chosen to optimize the increasing in the SP. 
In Fig.(\ref{fig:JanzskyModel}.b), we apply our method to plot  the SP as a function of time corresponding to such a frequency modulation. First, note that there is an increasing of $r(t)$ only when the frequency jumps  from $\omega_{0}$ to $\omega_{1}$, but not when the frequency jumps back from $\omega_1$ to the initial frequency $\omega_0$. In fact, after the frequency abruptly changes from $\omega_1$ to its original value $\omega_0$ the SP remains constant in time until the next jump from $\omega_{0}$ to $\omega_{1}$. This can be understood in the following way: 
after the jump from  $\omega_{0}$ to $\omega_{1}$ we showed that the state of the HO is a squeezed state of the original hamiltonian $\hat H_0$ (a HO with constant frequency  $\omega_{0}$), and it is known that the time-evolution  described by  $e^{-i\hat H_0 t/\hbar}$ of a squeezed state with respect to hamiltonian $\hat H_0$ does not change the value of the SP (though the variance of a quadrature operator oscillates with time due to its dependence on the squeezing phase $\varphi$). That is why in Fig.(\ref{fig:JanzskyModel}.b) we have plateaus whenever 
$\omega(t) = \omega_0$. Our description should be contrasted with that appearing in Refs.\cite{JANSZKY-TE-1994, JANSZKY-1992}, where it is suggested that the SP suffer abrupt (discontinuous) changes as the frequency jumps from $\omega_{0}$ and $\omega_{1}$ and from $\omega_{1}$ back to $\omega_{0}$. However, this is only an apparent disagreement since here the SP is always considered with respect to the original hamiltonian $\hat H_0$, while in the above mentioned papers, though not explicitly stated,  squeezing is considered  with respect to the instantaneous hamiltonian. 
It is worth mentioning that since finite changes in the HO frequency cause only finite changes in the corresponding hamiltonian,  the physical state of the HO evolves continuously in time since 
 \begin{equation}
 \lim_{\delta\rightarrow 0} e^{\frac{i}{\hbar}\hat H \delta} \vert\psi(t)\rangle = \vert\psi(t)\rangle\, .
 \end{equation}
 Hence, the same thing occurs with the SP, it can not suffer discontinuous changes, unless it is defined with respect to the instantaneous hamiltonian (which is not our case).
 In Ref.\cite{2019-AJP-Tiba}, we analyze a simplified version of the Janszky-Adam model which consists of one sudden frequency change from $\omega_{0}$ to $\omega_{1} > \omega_0$ (at $t=0$) followed by another sudden change from $\omega_1$ back to the initial frequency $\omega_{0}$ after  a time interval $\textsl{T}$. We obtain an exact analytical solution with the aid of algebraic methods based on Lie algebras and use this problem to unveil some qualitative aspects of squeezing processes by abrupt frequency changes. Particularly, we show why there is no change in the SP when the frequency jumps back to its original.

Note that, as in the parametric resonance case, the mean value of the SP grows linearly. As a consequence, it can be shown that both frequency modulations have a similar fingerprint in the complex space \textit{i.e.}, the curve exhibits a spiral like behavior, since the modulus of $z$ increases without bound, as in Fig.(\ref{fig:Resonance}.c).

\begin{figure}
\centering
\includegraphics*[width=15.0cm]{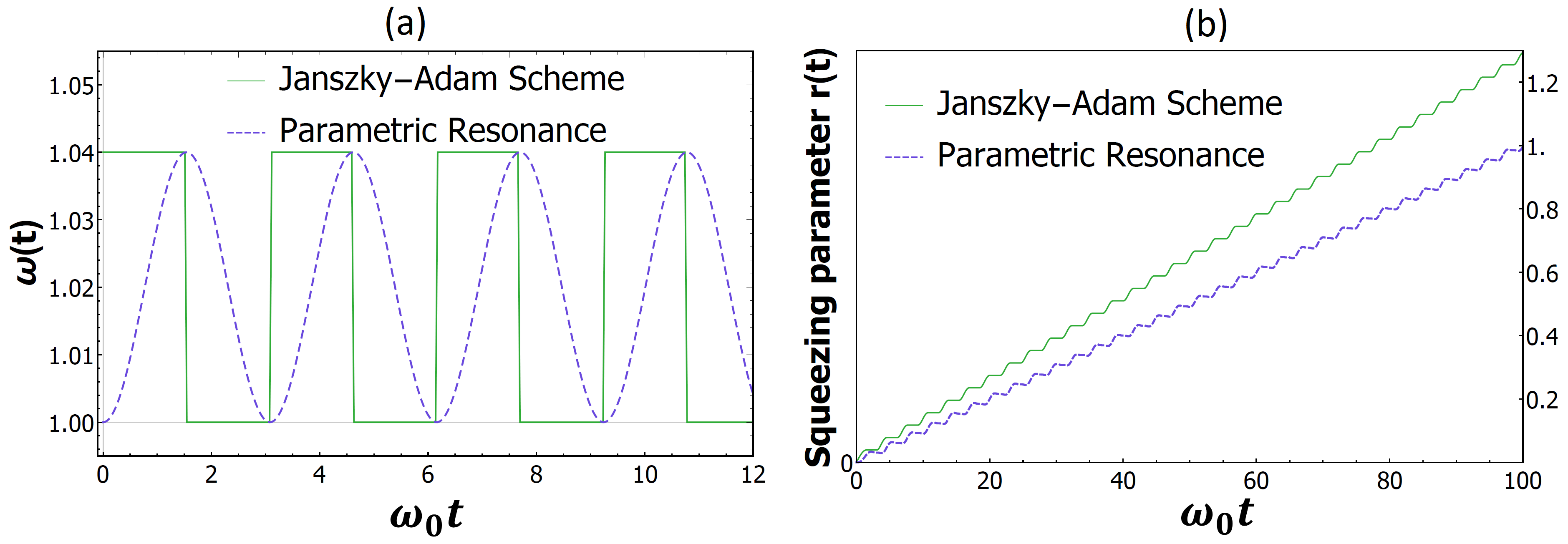}
\caption{(a) Plot of the frequency function for the J-A scheme and the parametric resonance model in time with the same minimum and maximum values, and (b) the resulting SP as a function of time.}
\label{fig:JanzskyModel2}
\end{figure}

Finally, in order to compare which process between the parametric resonance model and the Janszky-Adam scheme is more effective to squeeze the HO, we plot in Fig.(\ref{fig:JanzskyModel2}.a) the frequency modulations corresponding to these two models. Of course, for our comparison to make sense, we must choose appropriately the parameters in both models. Since the parametric resonance model to be used is that described by Eq. (\ref{eq:frequency3}), it is natural to choose both modulations between the same minimum and maximum frequency values, and  the amplitude is small, between 1.00 and 1.04 (in arbitrary units). In Fig.(\ref{fig:JanzskyModel2}.b) we plot the SP as a function of time for both models with the above choices for the parameters involved.  Although both curves have the same general form and show squeezing parameters that increase without bound, it is evident from Fig.(\ref{fig:JanzskyModel2}.b) that  the Janszky-Adam scheme is more efficient to squeeze than the parametric resonance (a similar conclusion was obtained by Galve and Lutz \cite{Galve-2009}).



\section{CONCLUSIONS} \label{C}

In this paper, using algebraic methods and appropriate BCH-like relations of Lie algebras we developed an iterative method for solving the problem of a harmonic oscillator with an arbitrary time-dependent frequency. Although the problem of a harmonic oscillator with a time-dependent frequency had already been  formally solved 
 by algebraic methods (see, for instance, Refs.\cite{Rhodes-1989, C.F.LO-1990}), our method has the advantage of being very well adapted for numerical calculations no matter the time dependence on the frequency. In other methods, only a few particular cases of time-dependent frequencies can be handled easily.  As it was already known in the literature, we have shown that a time-dependent frequency gives rise to a squeezed state. Our results enable us to follow the state of the system at any time and with the desired precision. As a consistence test, in order to get more confidence in our method, we first recovered some important results found in the literature \cite{JANSZKY-1994}. Then, we considered other important cases, namely, {\it (i)} the parametric resonance model and {\it (ii)} the Janszky-Adam scheme. 
By computing the squeezing parameter and the variances of quadrature operators for these models, we showed that the latter is the most efficient method for squeezing. 
 We think our method may be useful for a deeper understanding of squeezing procedures  as well as of general time-dependent problems involving  Lie algebras. 
Our method seems to be computationally attractive for the study of shortcuts to adiabaticity \cite{Chen-2010, Del-Campo-2011, guery-2019}, 
harmonic traps \cite{Grossmann-1995, schneiter-2020, qvarfort-2020}, 
and problems with coupled HO's \cite{urzua-2019, urzua-2-2019}
Moreover, since the HO with a time-dependent frequency appears in many different areas in physics, from quantum optics to quantum field theory in flat space-time (for instance in the dynamical Casimir effect) as well as in curved spacetimes (for instance in cosmological particle creation), we hope our method may inspire alternative ways of attacking problems of particle creation in general time-dependent backgrounds.


\section*{Acknowledgments}

The authors acknowledge R. Acosta Diaz, D. R. Herrera, L. Garcia, C. M. D. Solano, Reinaldo F. de Melo e Souza, M. V. Cougo-Pinto, A. Z. Khoury and P.A. Maia Neto for enlightening discussions. The authors thank the brazilian agencies  for scientific and technological research CAPES, CNPq and FAPERJ for partial  financial support.

\newpage
\bibliographystyle{unsrt}
\biboptions{sort&compress}
\bibliography{refs}

\end{document}